# LLMs for health: Perceived benefits, risks, intention to use AI chatbots, and willingness to self-disclose across sensitive health topics

*Gwenn Beets[1*], Anniek Jansen[1*], Saar Hommes[1,2], Ruben D. Vromans[1], Leonie Westerbeek[3], Supraja Sankaran[1], Julia C.M. van Weert[3], Emiel J. Krahmer[1,2], and Nadine Bol[1]*

[1] Department of Communication and Cognition, Tilburg School of Humanities and Digital Sciences, Tilburg University, Tilburg, The Netherlands

[2] Academic Collaborative Center for Digital Health and Mental Wellbeing, Department of Communication and Cognition, Tilburg School of Humanities and Digital Sciences, Tilburg University, Tilburg, The Netherlands

[3] Amsterdam School of Communication Research (ASCoR), Department of Communication Science, University of Amsterdam, Amsterdam, The Netherlands

* Authors contributed equally.

Corresponding author: Gwenn Beets, Department of Communication and Cognition, Tilburg University, P.O. Box 90153, 5000 LE Tilburg, g.beets@tilburguniversity.edu

## Abstract

AI chatbots are increasingly used for answering health-related questions. This study examines the role of topic type discussed with an AI chatbot and individual characteristics on perceived benefits and risks, intention to use an AI chatbot, and willingness to self-disclose health information. We conducted an online experiment with a 2 (topic type: physical versus psychological, between-subjects) × 2 (topic sensitivity: low versus high, within-subjects) mixed design among a Dutch representative sample ($N$ = 1,388). Results showed that perceived benefits were positively associated with intention and willingness to self-disclose, while perceived risks were negatively associated. Moreover, participants reported higher usage intentions for low-sensitive topics compared to high-sensitive topics. Furthermore, perceptions, intention, and willingness to self-disclose varied by individual characteristics. Overall, our findings suggest that intentions to use AI chatbots and self-disclosure of health-related information are primarily related to perceived benefits and risks and to personal characteristics rather than to topic type.

## 1 Introduction

People often worry about their mental and physical health, and may (increasingly) seek health-related information online (Xiao et al., 2024; Yun and Bickmore, 2025). With the recent rapid uptake of AI chatbots, chatbots based on Large Language Models (LLMs; i.e., language generation models trained on text), such as ChatGPT and DeepSeek, asking health-related questions online has begun to shift from static web searches to interactive dialogues (Xiao et al., 2024; Haensch, 2025; Yun and Bickmore, 2025). While search engines remain the primary source for health-related information, it is shown that AI chatbots are used by 37% of the Dutch population, according to a population-based study (de León et al., 2024). The increasing popularity of using AI chatbots for physical and psychological health-related information raises fundamental questions. Are chatbot users aware of benefits and risks? And if so, do these impact their intention to use chatbots and willingness to share personal sensitive health information with them? To what extent does this depend on the topic of discussion? In this paper, we tackle these questions using a large-scale online experiment among a representative Dutch sample.

Research has identified several benefits and risks to using AI chatbots for health-related questions (e.g., Singg and Pena, 2025). It remains unclear, however, how such perceived benefits and risks are associated with the intentions to use an AI chatbot and willingness to self-disclose personal information, and whether the topic of discussion plays a role for the general public. Research has suggested that people are more willing to accept AI chatbots and self-disclose information to an AI chatbot for stigmatized topics (Liu et al., 2025a; Liu et al., 2025b), perhaps since chatbots are engineered to be non-judgmental. It is unclear whether the topic type and sensitivity are associated with benefit and risk perceptions as well as willingness to use an AI chatbot and self-disclose. Therefore, we will systematically compare how differences in AI discussion topics shape public perceptions of benefits and risks and, ultimately their willingness to use and self-disclose to an AI chatbot.

Importantly, the general public is highly diverse, varying in characteristics such as gender, AI literacy, and political orientation. These individual differences are likely to be associated with intentions to use AI chatbots and willingness to self-disclose information. Although some studies explored the role of individual characteristics (Platt et al., 2024; Yun and Bickmore, 2025; Mendel et al., 2025), research remains limited. In addition to comparing the topic of discussion, we will provide a more nuanced view on perceptions, intention to use, and willingness to self-disclose by examining their associations with individual characteristics.

# 2 Theoretical framework

## 2.1 Trade-off between perceived benefits and risks

Underlying the intention to use AI chatbots and share personal information is a trade-off between perceived benefits and risks. People may outweigh expected benefits, such as having access to health and well-being services, with potential risks, such as exposure to misinformation, when deciding about their willingness to use AI chatbots (Tang et al., 2025). This trade-off is central to various theories and models, including the Unified Theory of Acceptance and Usage of Technology (UTUAT; Venkatesh et al., 2003), privacy calculus theory (Laufer and Wolfe, 1977), and Health Belief Model (HBM; Janz and Becker, 1984).

UTAUT describes how performance expectancy, effort expectancy, social influence, and facilitating conditions influence the intention to use a technology (Venkatesh et al., 2003). Applying UTAUT to AI chatbots, Menon and Shilpa (2023) showed how participants made a trade-off between perceived benefits of chatbots (e.g., being informative) and perceived risks (e.g., producing misinformation). The privacy calculus describes a similar trade-off by describing privacy concerns as influencing the willingness to self-disclose data. Furthermore, short-term benefits often outweigh long term-risks (Acquisti and Grossklags, 2005), which was also found in the context of AI chatbots where positive experiences were associated with more willingness to self-disclose while risks were negatively associated (Meng and Liu, 2025). From a health behaviour perspective, HBM describes how individual beliefs, including perceived benefits and risks, shape behaviour (Janz and Becker, 1984).

Overall, perceived benefits and risks are expected to play an important role in shaping the intention to use and willingness to self-disclose to an AI chatbot. We therefore expect that:

**H1**: Higher levels of perceived benefits are associated with (a) higher intentions to use an AI chatbot for health-related questions and (b) a higher willingness to self-disclose health-related information to an AI chatbot.

**H2**: Higher levels of perceived risks are associated with (a) lower intentions to use an AI chatbot for health-related questions and (b) a lower willingness to self-disclose health-related information to an AI chatbot.

## 2.2 Discussing different health topics

Given the complex organizational structures and fragmented nature of healthcare amongst others, people are turning to AI chatbots with health-related questions concerning different types of health topics (physical or psychological) they are navigating, with varying levels of sensitivity (Costa-Gomes et al., 2026). This fragmentation is also visible in research, where studies oftentimes focus on a single healthcare context. For example, AI chatbots have been studied regarding discussions about lifestyle behaviours (Singh et al., 2023), empowering cancer patients (Shaban et al., 2025), and improving mental health in college students (Nyakhar and Wang, 2025). Regardless of the topic type being discussed with an AI chatbot, topic sensitivity may play a crucial role. Topic sensitivity refers to the extent to which complex, often emotionally charged or socially fraught issues may be difficult to discuss openly due to, for instance, stigma or cultural norms (adapted from Macleod et al., 2025). Cultural norms and normative expectations have been shown to shape disclosure behaviour, as they form the basis for the judgement of appropriateness to self-disclose in a certain context (Nissenbaum, 2004). Hence, self-disclosure is context-dependent, more likely to happen in situations that are deemed less intimate (Taylor and Altman, 1975), and where self-disclosure has high perceived utility and low perceived risks (Omarzu, 2000).

Despite the growing use of AI chatbots for health communication, the role of discussion topics has received little research attention. Although AI chatbots were generally perceived as the least acceptable source for medical consultations, they were considered more acceptable for stigmatised topics compared to low stigmatised topics (Miles et al., 2021). Moreover, perceived health stigma was found to be positively related to willingness to self-disclose information to an AI chatbot (Liu et al., 2025b). Findings regarding benefits, risks, and use are fragmented and not systematically explored in relation to topic type and sensitivity. Therefore, we formulated the following research question (RQ):

**RQ1**: What are the effects of topic type (physical versus psychological conditions) and topic sensitivity (low- versus high-sensitive) on (a) perceived benefits, (b) perceived risks, (c) intention to use, and (d) willingness to self-disclose for health-related questions?

## 2.3 Individual differences in using and perceiving AI chatbots

Individual differences can also shape perceptions and usage. Jia et al. (2021) found that individual characteristics, such as age, gender, literacy, and education, were related to online health-information seeking. Additionally, individual differences and experience with the technology, have been related to technology acceptance and use (Yi et al., 2005). Sociodemographic and individual differences may also reflect existing socioeconomic structures and digital inequalities in society (Lythreatis et al., 2022). For instance, previous work on health-related AI chatbots has shown that females expected fewer benefits from AI chatbots, were less likely to use them (Platt et al., 2024; Mendel et al., 2025), and perceived AI chatbots for mental health as less beneficial for the future (Benda et al., 2024). Younger age was furthermore shown to be associated with less expected health benefits from using an AI chatbot (Platt et al., 2024), while Yun and Bickmore (2025) reported that older age was correlated with lower use of AI chatbots for health information. Additionally, people were more likely to use AI chatbots for their mental health if they had also spoken about it with a physician prior (McBain et al., 2026).

Overall, various factors can influence how individuals perceive and use AI chatbots. We address these in RQ2:

**RQ2:** How are (a) perceived benefits, (b) perceived risks, (c) intention to use AI chatbots for health-related questions, and (d) willingness to self-disclose related to individual factors (i.e., age, gender, AI literacy, eHealth literacy, education, voting behaviour, trust in healthcare institutions, trust in institutions, general health, use of AI chatbots as a health information source, monitoring coping style, AI usage experience, frequency of visiting health services)?

# 3 Methods

The study was approved by the Research Ethics and Data Management Committee of Tilburg University (2025.46) prior to data collection. The study was preregistered within the Open Science Framework (OSF) prior to accessing the data. A description of the deviations from the preregistration can be found on OSF Appendix A as well as related materials and code (https://osf.io/njbp5/overview?view_only=7083664909124b4d8389424afd8681d7) .

## 3.1 Design

We conducted an online experiment with a 2 (topic type: physical versus psychological condition; between-subjects) × 2 (topic sensitivity: low versus high; within-subjects) mixed factorial design, resulting in sixteen scenarios across four experimental conditions, of which each participant saw two.

## 3.2 Participants and power analysis

Participants were recruited through the LISS panel (Longitudinal Internet studies for the Social Sciences), managed by the non-profit research institute Centerdata (Tilburg, the Netherlands). The panel is based on a true probability sample of households drawn from the population register by Statistics Netherlands. Panel members could participate if they (1) were 18 years or older, (2) participated in the third wave of the AlgoSoc study (a prior study on the topic of AI), and (3) provided informed consent. Participants completed the questionnaire in September 2025 and received three euros upon completion following the panel standards.

An a-priori power analysis was conducted through G*Power to determine the minimally required sample size for detecting small effects. To obtain 95% power to detect an effect size of .10 with a .05 alpha error probability based on a MANOVA (RQ1), we required a sample of 978 participants. We chose to oversample to identify individual variations in perceptions (RQ2) and account for missing data on background characteristics from existing datasets, thus aiming for a minimum sample size of 1,200 participants.

## 3.3 Stimuli

To cover a broad range of health conditions and ensure this study's findings reflect the broader experimental conditions of topic type and sensitivity, we created sixteen scenarios. For each experimental condition, two health-related conditions were selected from 24 health conditions based on a pretest resulting in eight conditions. These were presented in two scenarios describing two types of interactions; either a user chatting with an AI chatbot or asking for a summary (see Figure 1).

**Example scenario A: Physical, low-sensitive condition | Chat interaction**
Imagine you have anemia. You recently visited a doctor for this. After the consultation, you still have some questions about your symptoms and how best to manage your anemia.
You would like answers to your questions. That is why you start a chat with an AI chatbot. In this chat, you ask questions about your symptoms and how to manage your anemia. A conversation follows in which the AI chatbot asks follow-up questions and then answers all your questions. You then end the conversation with the AI chatbot.

**Example scenario B: Psychological, high-sensitive condition | Summary interaction**
Imagine you have an eating disorder. You recently visited a psychologist for this. After the consultation, you will have received a letter with information on how best to deal with your eating disorder.
The letter is very long and therefore you would like to have a summary of this letter. You take a photo of the letter and send it to an AI chatbot. The AI chatbot will summarize the letter at your request for you. Then read the summary carefully. After that, you end the interaction with the AI chatbot.

**Variations related to the experimental conditions**
Health condition
*Based on topic type and topic sensitivity, selected from 8 medical conditions.*
Healthcare provider
*Based on topic type.*

**Figure 1** Examples of the stimuli scenarios.

## 3.1 Pretest

Forty-four participants completed a pretest. Participants were recruited via the researchers' networks. Twenty-four health conditions (twelve physical, twelve psychological) were preselected from a broad list of common health conditions through careful discussion among the authors (OSF Appendix B&C). Participants rated all health conditions on topic sensitivity and severity. Topic sensitivity was measured by asking participants to answer the question "To what extent would you describe the health condition as sensitive to discuss?" on a 5-point Likert scale (1 = '*Not at all sensitive'* to 5 = *'Very sensitive'*). Topic severity was evaluated by answering the question 'To what extent would you describe the condition as mild or severe (in terms of severity)?' on a 5-point Likert scale (1 = *'Very mild*' to 5 = '*Very severe'*).

We aimed to select conditions that were equal in sensitivity within the experimental condition (e.g., low-sensitive physical and psychological) and significantly different from the other experimental conditions (e.g., high-sensitive physical and psychological). It was not possible to select equally sensitive topics for psychological and physical conditions, since psychological conditions were generally perceived as more sensitive compared to physical conditions. Ultimately, we selected health conditions that differed in topic sensitivity within the topic type category and scored comparable on perceived severity. A detailed description of the results can be found on OSF in Appendix D.

## 3.2 Procedure

After informed consent, participants answered questions about their personal characteristics such as coping style, eHealth, AI literacy, and their experience with AI chatbots. Next, they received an explanation of AI chatbots to ensure a shared understanding. They were then asked to list their perceived benefits and risks of using AI chatbots for health. Afterwards, participants were presented with two scenarios describing the use of an AI chatbot for a health condition. Both scenarios per participant described either physical or psychological conditions but varied in sensitivity level (low or high) and type of chatbot interaction (chat or summary), in random order. After each scenario, participants answered manipulation check questions indicating their experience with the described health condition, and answered questions about their intention to use, willingness to self-disclose health-related information, and perceived benefits and risks of using an AI chatbot in the scenario. Upon completion of the questionnaire, participants were debriefed about the study goal and provided with additional resources related to AI use and physical and psychological health. A visual overview of the procedure can be found in Figure 2 and the full questionnaire on OSF Appendix E&F.

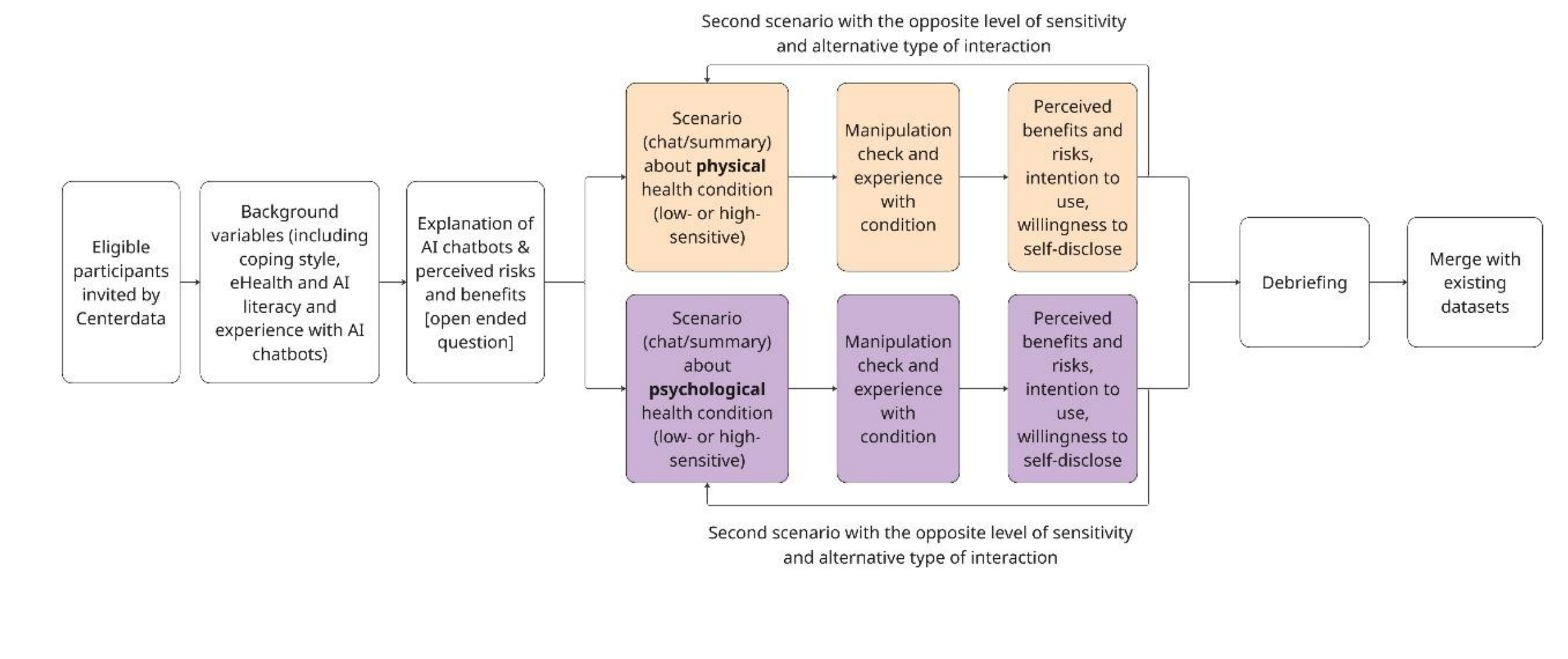

**Figure 2** Visual overview of the study procedure.

## 3.3 Measures

Variables were either measured in the current questionnaire (♣) or extracted from previously collected LISS datasets: Background variables (♦), Health (●), Politics and Values (■), and the third wave of the AlgoSoc Longitudinal Survey Panel study (de León et al., 2024) (▲).

### 3.3.1 Dependent variables

Measurement invariance was assessed and confirmed for all dependent variables (see OSF Appendix G). Additionally, exploratory factor analyses were conducted for both benefits and risks scales (OSF Appendix H).

#### *3.3.1.1 Perceived benefits ♣*

Perceived benefits were measured by asking participants to rate the extent to which they see certain aspects as a benefit of AI chatbots (e.g., "It is a benefit that the AI chatbot… can answer an unlimited number of questions"), on a 5-point Likert scale ranging from 1 '*Totally disagree*' to 5 '*Totally agree*'. Statements were adapted from a taxonomy by Antes et al. (2021). Average benefit scores were calculated for low- and high-sensitive scenarios, with excellent internal consistency (Cronbach's α = .93 and .94, respectively).

#### *3.3.1.2 Perceived risks ♣*

Perceived risks were assessed by asking participants how much they worry about certain risks of using AI chatbots. Ten risks were presented (e.g., "It is a risk that the AI chatbot …could leak private information") and assessed on a 5-point Likert scale ranging from 1 '*Totally disagree*' to 5 '*Totally agree*'. Statements were derived from a taxonomy by Weidinger et al. (2022). The ten items were averaged into a score for low- and high-sensitive scenarios, with good internal consistency (Cronbach's α = .87 and .86, respectively).

#### *3.3.1.3 Intention to use AI chatbots ♣*

Adapted from a previous study (Reeskens and Metz, 2023), intention to use an AI chatbot in the scenario was assessed using three items (e.g., "In the described scenario, I would be willing to use this AI chatbot.") on a 5-point Likert scale ranging from 1 '*Completely disagree*' to 5 '*Completely agree*'. For both scenarios, the three items were averaged into a scale to create risk-scores (Cronbach's $\alpha_{low}$= .95; Cronbach's $\alpha_{high}$ = .95).

#### *3.3.1.4 Willingness to self-disclose* ♣

Willingness to disclose information about oneself to an AI chatbot was assessed using three items. Participants answered the following question: "How likely is it that you would want to share the following personal information with the AI chatbot in the described scenario?" for three items (e.g., "your medical data"), on a 5-point scale ranging from 1 '*Very unlikely*' to 5 '*Very likely*'. The three items were averaged into a score for low- and high-sensitive scenarios, with good internal consistency (Cronbach's α = .87 and .87, respectively).

### 3.3.2 Personal characteristics

#### *3.3.2.1 Monitoring coping style* ♣

The extent to which participants use a monitoring coping style was assessed using three items from the 'Threatening medical situations inventory' (e.g., "I intend to ask the specialist as many questions as possible."; Van Zuuren et al., 1996), answered on a 5-point scale (1 '*Not at all applicable to me*' to 5 '*Very much applicable to me*'). We used one of the inventory's scenarios (i.e., experiencing a headache). The internal consistency of the three items was poor (Chronbach's α = .48) and could not be improved by removing one of the items. Therefore, the items were included separately in the analyses.

#### *3.3.2.2 AI literacy* ▲♣

AI literacy was measured using four dimensions: Know & Understand, Use & Apply, AI ethics and Detect AI (Long and Magerko, 2020; Wang et al., 2022). Data on the first three dimensions were previously collected by the AlgoSoc study, the latter dimension was measured in our questionnaire. To assess the level of AI literacy, participants were asked to answer the following question: 'How good do you think you are at the following tasks?' for items related to each dimension (e.g., "Recognizing whether I am dealing with an application that uses AI"). Participants answered using a slider ranging from 0 '*Not good at all*' to 10 '*Very good*'. The eighteen items were averaged into a score for AI literacy, with excellent internal consistency (Chronbach's α = .97).

#### *3.3.2.3 eHealth literacy* ♣

We used eight items to measure eHealth literacy (e.g., "I know which health information can be found on the internet") on a 5-point Likert scale ranging from 1 '*Very much disagree*' to 5 '*Very much agree*' (Norman and Skinner, 2006; Van Der Vaart et al., 2011). The average score of all combined items had excellent internal consistency (Chronbach's α = .94).

#### *3.3.2.4 AI chatbot experience* ♣

We assessed whether participants had heard of AI chatbots with one question: "Have you ever heard of an AI chatbot?" ('*Yes*', '*No*'). Participants who indicated to have heard of AI chatbots, were asked the follow-up question: "Have you ever used an AI chatbot?" ('*Yes*', '*No*', '*I don't know*'). Answers to these two questions were combined into one variable 'AI chatbot experience' with the answer options '*Not heard of, not used', 'Heard of, but not used', 'Heard of, use unknown',* and *'Heard of and used'.*

#### *3.3.2.5 Personal experience with health conditions* ♣

Participants were asked whether they have (had) the health condition described in the scenario they read: "Do you have [insert health condition] at the moment or have you ever had it?" (*'Yes', 'No', 'Prefer not to say'*).

#### *3.3.2.6 Use of chatbot for health information* ▲

Chatbot use for health information was assessed using one item of a longer scale. Participants answered the following question: "How often have you used the following sources for health-related purposes?" for the item 'chatbot or virtual assistants (such as ChatGPT, Bard/Gemini, Bing Chat)'.

They answered this on a 7-point scale, ranging from 1 '*Never'* to 7 '*Always for a health-related question*'.

#### *3.3.2.7 Frequency of using health services ●*

Frequency of using four different health services ('Psychiatrist/psychologist/psychotherapist', 'General practitioner', 'Physiotherapist', 'Medical specialist') was assessed by asking participants to indicate how many times in the past twelve months they had visited each of these services.

#### *3.3.2.8 General health ●*

General health was assessed using one question where participants rated their health in general on a 5-point scale, ranging from 1 '*Bad'* to 5 '*Excellent'*.

#### *3.3.2.9 Trust in institutions ■*

Trust in institutions was assessed by asking participants to indicate how much trust they had in eight different organizations (e.g., the army, healthcare, democracy), on a scale from 1 '*No trust at all*' to 10 '*Complete trust*'. Trust in healthcare was included separately in the analyses. The remaining seven items were averaged into one score measuring trust in other institutions (Cronbach's α = .88).

#### *3.3.2.10 Political orientation ■*

Participants were asked for which political party they voted during the Dutch national elections in 2023, with answer options for all Dutch parties participating in the 2023 elections and for blank, non-disclosed, and non-voted. The data was recoded to right- or left-wing political orientation or other based on the political landscape at that time (Kieskompas, 2023).

#### *3.3.2.11 Demographic characteristics ♦*

Age was measured with an open answer indicating the participant's age in years. Gender was measured categorically: '*Male'*, '*Female'*, and '*Other'*. Obtained education level was assessed following the Dutch education system ('*Primary school*', '*Pre-vocational education*', '*Secondary education*', '*Post-secondary vocational education*', '*Higher professional education*', '*University*', '*Other'*) and aggregated into 'low', 'middle', and 'high' education levels based on existing schemes (Statistics Netherlands: no date).

### 3.3.3 Manipulation check ♣

A manipulation check assessed whether the conditions were perceived as intended in terms of sensitivity. Participants were asked to answer: "To what extent would you describe [insert condition] as sensitive to discuss?" on a 5-point Likert scale ranging from 1 '*Not at all sensitive*' to 5 '*Very sensitive*'. Participants were also asked to answer the question 'To what extent would you describe the condition as mild or severe (in terms of severity)?' on a 5-point Likert scale ranging from 1 '*Very mild*' to 5 '*Very severe'.* Severity was included to account for comparable severity across conditions. However, it became clear in the pretest that it was not possible to select health-conditions which were not significantly different in severity, however they were comparable.

## 3.4 Statistical analyses

All pre-registered analyses were performed using R Statistical Software (v4.5.2.,R Core Team, 2016) using the *lme4*, *ggplot2*, and *sjPlot* packages. To answer RQ1, a mixed MANOVA was conducted to evaluate the effect of topic type and sensitivity on the perceived benefits, perceived risks, intention to use, and willingness to self-disclose. Four separate ANOVAs were performed as post-hoc tests to check for effects of topic type and sensitivity on the four DVs.

To test H1 and H2, separate linear mixed effects models were fitted for intention to use and willingness to self-disclose in a stepwise approach. First, a baseline model (with random intercept for participants only) was fitted to account for individual differences in baseline scores on the outcomes.

Second, we fitted separate fixed-effects models: one with benefits and one with risks as fixed effects and random intercept for participants. Third, an extended random intercepts model was fitted in which random intercepts were added for topic type, topic sensitivity, and their interaction effect to account for differing baseline scores for these experimental conditions.[1]

We compared their model fit using an ANOVA, selecting the model that was a significant improvement compared to the previous model.

Two additional models were fitted for intention to use and willingness to self-disclose with perceived benefits and risks combined into one model.

To answer RQ2, a similar approach was adopted to assess the relation between personal characteristics and all four dependent variables. The fixed effects in Step 2 included: gender, age, AI literacy, general health, political behaviour, AI chatbot experience, use chatbot for health information, overall trust in institutions, trust in healthcare, education, frequency of visiting medical specialists, GP, physiotherapists, psychologist, eHealth literacy, coping monitoring as individual items. The models for intention to use and willingness to self-disclose also included perceived benefits and risks as fixed effects.

For all models, except the model with perceived benefits and intention to use (H1), the fixed effects model (Step 2) was the best fitting model. For consistency, we presented the results of the fixed effect models for all analyses. For the hypotheses and RQ2, we reported standardised estimates by entering standardised variables into the linear mixed effects model.

To account for multiple-testing an alpha-correction was applied. For H1 and H2, the critical *p*-value was set to .025 as it required two tests and for RQ1 and RQ2 to .0125 as it required four models to answer.

# 4 Results

Below, a description of all results related to the hypotheses and research questions can be found. Extended tables can be found in the supplementary materials.

## 4.1 Participants

A total of 1,720 respondents were invited by Centerdata to ensure a minimum of 1200 responses, 1,394 completed the full questionnaire. Six participants who were younger than 18 were removed, resulting in 1,388 eligible participants (response rate: 80.7%). An overview of the distribution of participants across conditions is in Figure 3. Participants were on average 51.4 years old (SD = 18.4, range = 18–97), with 695 being female (50.1%), and almost half being highly educated (n = 652, 47.0%). A detailed description of participant characteristics is in Table 1.

[1] A fourth step of fitting a random slopes model which included random slopes topic type*topic sensitivity was also fitted. These models all returned an error indicating that the number of observations <= number of random effects and was therefore not considered for best fitting model.

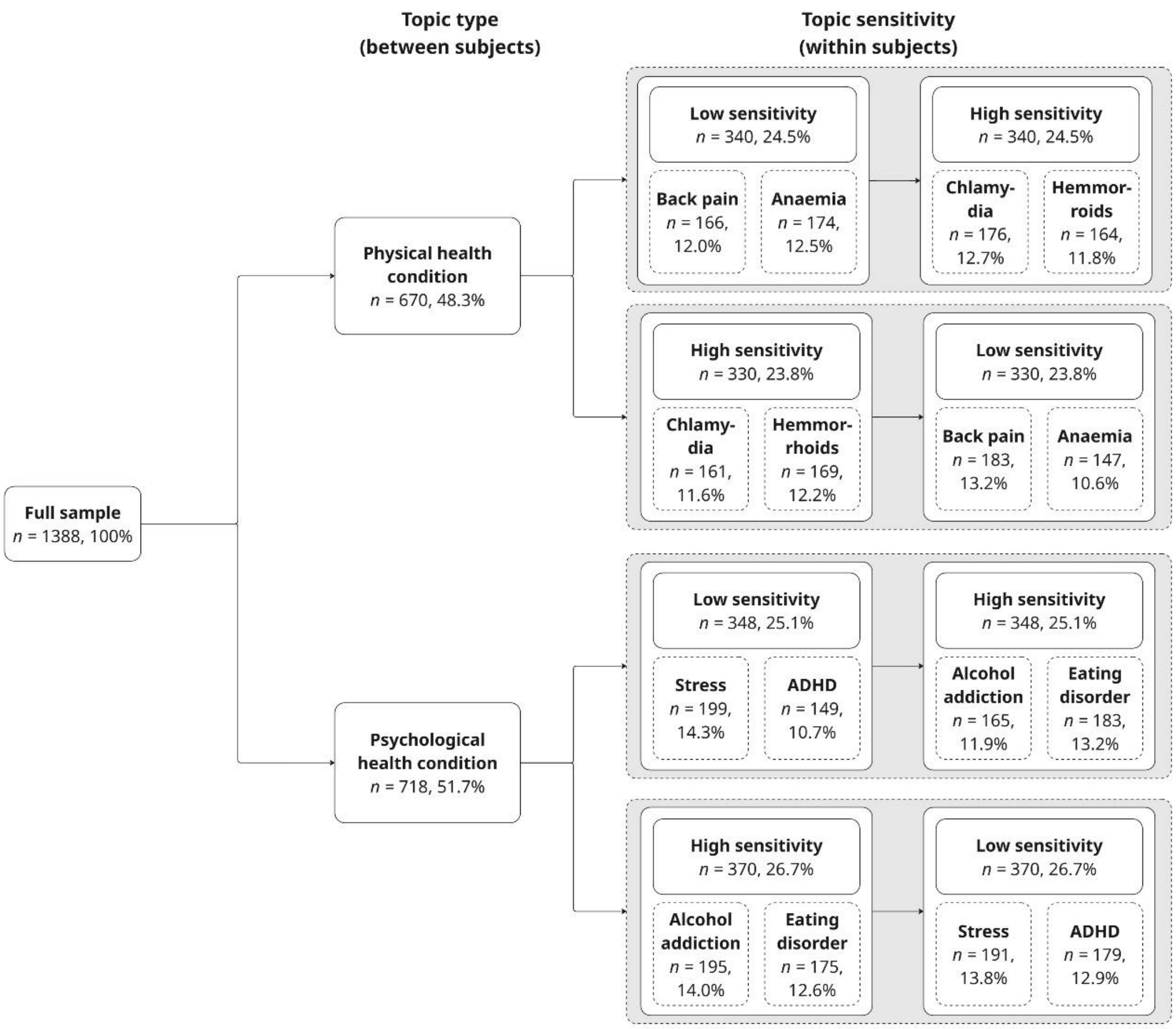

**Figure 3** Overview of the distribution of participants across experimental conditions.

**Table 1** Participants' characteristics compared to the Dutch population.

| Participant Characteristics | *n* (%) | Dutch population *n* (x1000) (%)[1] |
|---|---|---|
| **Age** *M* (*SD*) | 51.37 (18.36) | |
| 18-24 | 118 (8.5%) | 1,599 (10.8%) |
| 25-34 | 197 (14.2%) | 2,422 (16.3%) |
| 35-44 | 195 (14.0%) | 2,287 (15.4%) |
| 45-54 | 237 (17.1%) | 2,206 (14.9%) |
| 55-64 | 253 (18.2%) | 2,506 (16.9%) |
| 65+ | 388 (28.0%) | 3,834 (25.8%) |
| **Gender** | | |
| Women | 695 (50.1%) | 7,465 (51.0%) |
| Men | 690 (49.7%) | 7,182 (49.0%) |
| Non-binary | 3 (0.2%) | - |
| **Education**[2] | | |
| Low | 217 (15.6%) | 4,082 (27.1%) |
| Middle | 477 (34.4%) | 5,274 (35.0%) |
| High | 652 (47.0%) | 5,600 (37.2%) |
| Other (Other, no education, no information) | 42 (3.0%) | 94 (0.6%) |
| **Political orientation** | | |
| Left | 587 (42.3%) | 4,844 (46.2%) |
| Right | 466 (33.6%) | 5,393 (51.5%) |
| Other (No information/not mentioned/other parties) | 335 (24.1%)[3] | 238 (2.3%) |
| **Heard/used AI chatbots** | | |
| Not heard of | 147 (10.6%) | |
| Heard of, but not used | 389 (28.0%) | |
| Heard of, and used | 786 (56.6%) | |
| Heard of, and use unknown | 66 (4.8%) | |
| **eHealth literacy** *M* (*SD*) | 3.69 (0.73) | |
| **AI literacy** *M* (*SD*) | 4.14 (2.13) | |

| | |
|---|---|
| **Monitoring coping style aspects** *M* (*SD*) | |
| Ask questions to specialist | 3.74 (1.00) |
| Check other institutions or doctors | 2.14 (0.94) |
| Search for more information | 3.39 (1.13 |
| **Number of visits to health institutions in last 12 months** *M* (*SD*) | |
| General practitioner | 1.75 (2.74) |
| Medical specialist | 1.11 (3.17) |
| Physiotherapist | 2.64 (8.70) |
| Psychiatrist/psychologist/psychotherapist | 1.05 (5.59) |
| **General health** *M* (*SD*) | 3.16 (0.79) |
| **Trust in institutions (overall)** *M* (*SD*) | 6.53 (1.32) |
| **Trust in healthcare** *M* (*SD*) | 7.15 (1.64) |
| **Use of chatbot for health information** *M* (*SD*) | 1.53 (1.22) |

Note. [1] Data regarding the Dutch population in 2025/2025 from Statistics Netherlands https://www.cbs.nl/en-gb.
[2]The data for education are based on the sample of all people of 15 years and older.
[3]A relatively large proportion did not disclose their voting behaviour, explaining the large group of 'other'.

## 4.2 Manipulation check

Pairwise comparisons showed significant differences between all four experimental conditions on sensitivity ratings (all comparisons: $p < .001$). High-sensitive physical ($M = 3.65$, $SD = 0.98$) conditions were rated significantly more sensitive than low-sensitive physical conditions ($M = 2.22$, $SD = 0.88$). This also held for the psychological conditions ($M_{high} = 3.91$, $SD = 0.84$, $M_{low} = 3.16$, $SD = 0.91$). Overall, the psychological conditions were rated as more sensitive than physical conditions (see for more details OSF Appendix H). The manipulation check was therefore partially successful.

## 4.3 H1 & H2: Perceived benefits and risks

Overall, participants perceived relatively high levels of benefits ($M = 3.65$, $SD = 0.73$) and risks ($M = 3.54$, $SD = 0.61$), and scored around the midpoint in terms of intention to use ($M = 2.79$, $SD = 1.05$) and willingness to self-disclose ($M = 2.61$, $SD = 1.01$) (Figure 4). Greater perceived benefits were associated with a higher intention to use ($b = 0.51$, $SE = 0.02$, $p < .001$), whereas greater perceived risks were associated with a lower intention to use AI chatbots ($b = -0.14$, $SE = 0.02$, $p < .001$). These effects also held in the model with both perceived benefits and risks ($b_{benefits} = 0.51$, $SE_{benefits} = 0.02$, $p < .001$; $b_{risks} = -0.16$, $SE_{risks} = 0.02$, $p < .001$), confirming H1a and H2a. Likewise, greater perceived benefits were associated with a higher willingness to self-disclose ($b = 0.45$, $SE = 0.02$, $p < .001$), while greater perceived risks were associated with a lower willingness to self-disclose ($b = -0.15$, $SE = 0.02$, $p < .001$), also when considering perceived benefits and risks simultaneously (benefits: $b = 0.46$, $SE = 0.02$, $p < .001$; risks: $b = -0.18$, $SE = 0.02$, $p < .001$). These findings confirmed H1b and H2b.

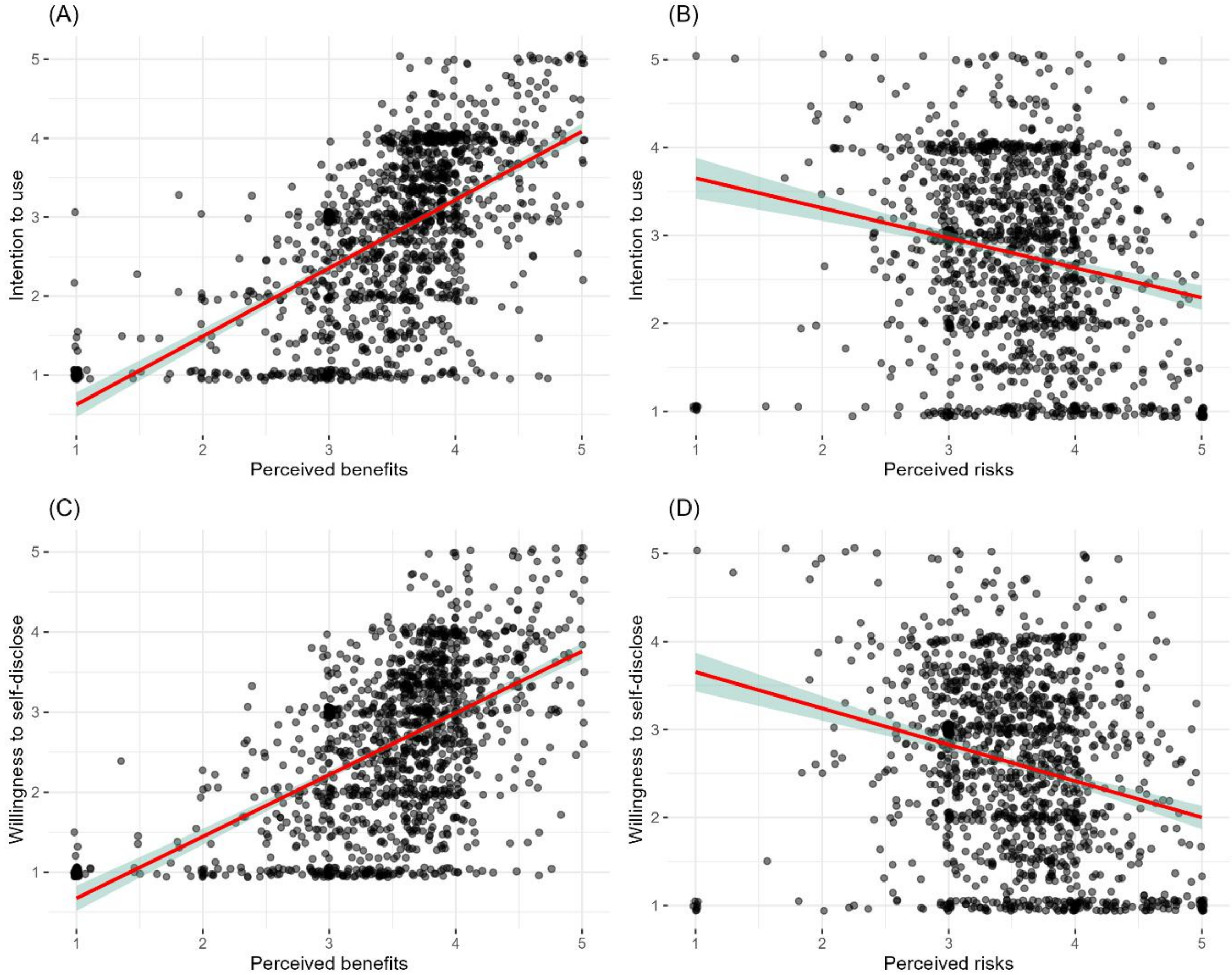

**Figure 4** Scatter plots for the association between intention to use and perceived benefits (A) and risks (B) and willingness to self-disclose and perceived benefits (C) and risks (D)

## 4.4 RQ1: Topic characteristics

While we found a multivariate effect of topic type on perceived benefits and risk, intention to use, and willingness to self-disclose (Pillai = .006, $F(4, 2767) = 4.06$, $p = .003$), we only found one univariate effect of topic sensitivity on intention to use the AI chatbot, $F(1, 1386) = 11.74$, $p < .001$, eta-squared = .001. Effect sizes were small, and differences across topic type and sensitivity were, in general, negligible (see Figure 5). Intention to use an AI chatbot was slightly higher in low-sensitivity conditions ($M = 2.83$, $SD = 1.13$) than in high-sensitive conditions ($M = 2.75$, $SD = 1.14$).

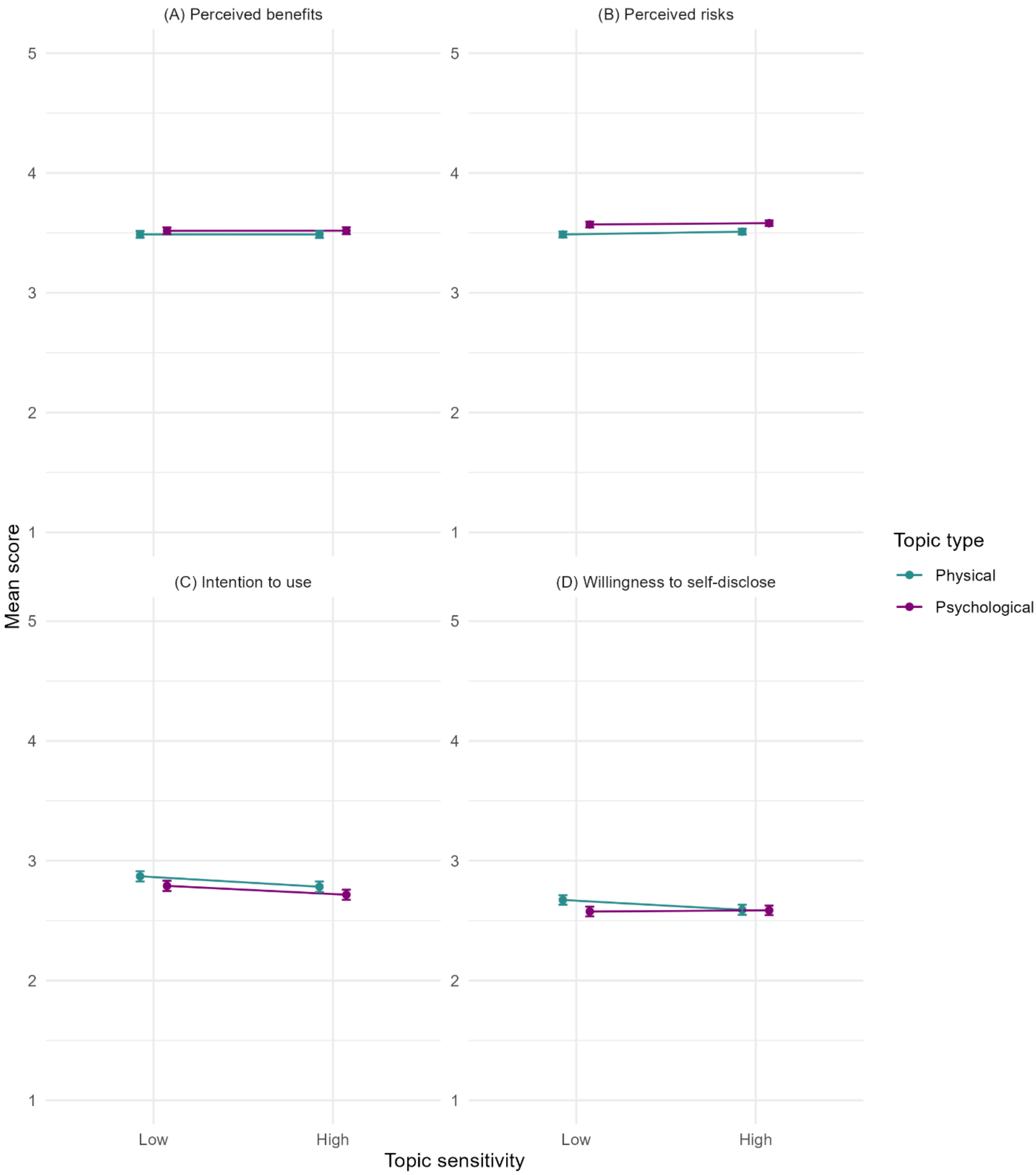

**Figure 5** Results of four ANOVAs examining the effects of topic type and sensitivity on perceived benefits and risks, intention to use and willingness to self-disclose.

## 4.5 RQ2: Influence of individual factors

The results of the fixed effects models are in Figure 6, visualised and the significant associations are described below.

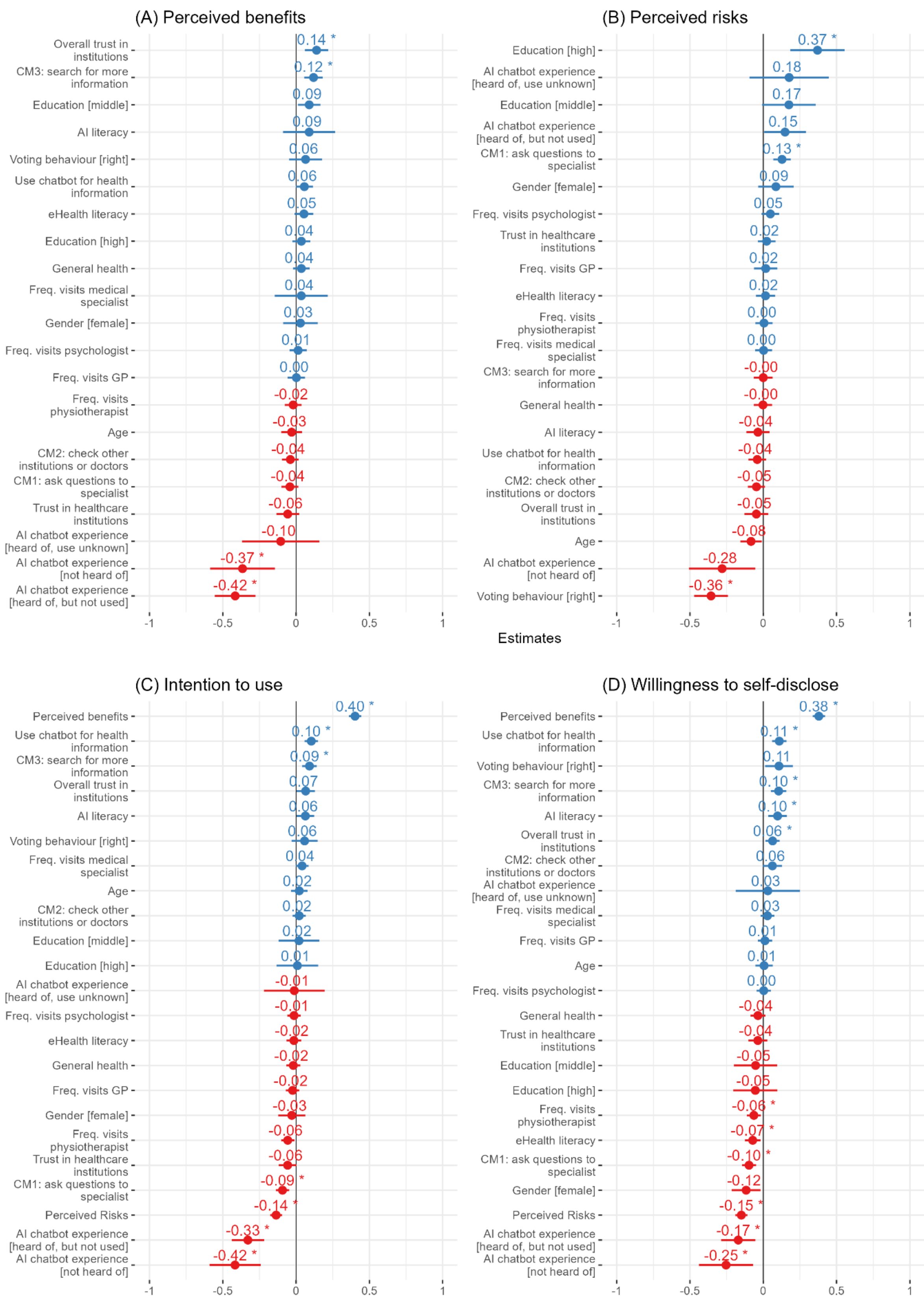

**Figure 6** Forest plot showing the associations between individual characteristics and perceived benefits and risks, intention to use, and willingness to self-disclose. * $p < .0125$. Reference categories are Gender [male], Voting behaviour [left], AI chatbot experience [heard of and used], Education [low].

#### 4.5.1 Perceived benefits

Compared to those who had heard of and used AI chatbots before, participants perceived lesser benefits if they had heard of AI chatbots but not used them ($b$ = -0.42, $SE$ = 0.07, $p$ < .001) or did not hear of them before ($b$ = -0.37, $SE$ = 0.11, $p$ = .001). Additionally, trust in institutions was positively associated with perceived benefits ($b$ = 0.14, $SE$ = 0.04, $p$ < .001), indicating that more trust was associated with greater perceived benefits. A higher intention to retrieve information about a condition, an aspect of a monitoring coping style, was associated with greater perceived benefits ($b$ = 0.12, $SE$ = 0.03, $p$ < .001).

#### 4.5.2 Perceived risks

Participants with a higher education level perceived greater risks compared to those with a lower education level ($b$ = 0.37, $SE$ = 0.09, $p$ < .001). Moreover, participants with a left political orientation perceived greater risks than those with a right political orientation ($b$ = -0.36, $SE$ = 0.06, $p$ < .001). Additionally, a higher intention to ask a specialist in the hospital as many questions as possible – another aspect of monitoring coping style – was associated with greater risk perceptions ($b$ = 0.13, $SE$ = 0.03, $p$ < .001).

#### 4.5.3 Intention to use

More frequent use of an AI chatbot for health-related information was related to higher intentions to use an AI chatbot in the future ($b$ = 0.10, $SE$ = 0.02, $p$ < .001). Participants who had neither heard of, nor used AI chatbots had a lower intention, compared to participants who had used AI chatbots before ($b$ = -0.42, $SE$ = 0.09, $p$ < .001). Similar patterns emerged for those who heard of AI chatbots but had not used them before ($b$ = -0.33, $SE$ = 0.06, $p$ < .001). Moreover, the more participants intended to ask a medical specialist in the hospital questions, the lower their intention to use an AI chatbot ($b$ = -0.09, $SE$ = 0.02, $p$ < .001), whereas the more participants intended to search for information themselves, the higher their intention to use ($b$ = 0.09, $SE$ = 0.03, $p$ < .001). Lastly, greater perceptions of benefits were associated with higher intention to use ($b$ = 0.40, $SE$ = 0.02, $p$ < .001), whereas greater perceptions of risks were associated with a lower intention to use ($b$ = -0.14, $SE$ = 0.02, $p$ < .001).

#### 4.5.4 Willingness to self-disclose

Higher AI literacy ($b$ = 0.10, $SE$ = 0.03, $p$ = .003) and frequency of using an AI chatbot for health information ($b$ = 0.11, $SE$ = 0.03, $p$ < .001) were associated with a higher willingness to self-disclose. On the contrary, higher eHealth literacy ($b$ = -0.07, $SE$ = 0.03, $p$ = .008) and frequency of visiting the physiotherapist ($b$ = -0.64, $SE$ = 0.02, $p$ = .009) were associated with a lower willingness to self-disclose. Additionally, participants who had neither heard of nor used AI chatbots had a lower willingness to self-disclose, compared to participants who used AI chatbots before ($b$ = -0.25, $SE$ = 0.09, $p$ = .007). Similar patterns emerged for those who heard of AI chatbots, but had not used them before ($b$ = -0.17, $SE$ = 0.06, $p$ = .004). Furthermore, the more participants intended to ask a medical specialist in the hospital more questions, the lower their willingness to self-disclose ($b$ = -0.10, $SE$ = 0.02, $p$ < .001). Participants had a higher willingness to self-disclose when they decided to inform themselves through other institutions and doctors ($b$ = 0.06, $SE$ = 0 .02, $p$ = .009) or intended to search for more information about the condition ($b$ = 0.10, $SE$ = 0.03, $p$ < .001). Lastly, greater perceptions of benefits were associated with greater willingness to self-disclose ($b$ = 0.38, $SE$ = 0.02, $p$ < .001), whereas greater perceptions of risks were associated with lower willingness to self-disclose ($b$ = -0.15, $SE$ = 0.02, $p$ < .001).

# 5 Discussion

We conducted an online experiment among a representative Dutch sample to systematically investigate the role of the discussion topic and individual characteristics on the perceived benefits and risks, intention to use AI chatbots, and willingness to self-disclose.

As predicted, our analysis showed that greater perceived benefits are associated with higher intention to use and self-disclose, while greater perceived risks are associated with lower intention to use and self-disclose (confirming H1 & H2). These findings align with HBM (Janz and Becker, 1984) and the privacy calculus (Laufer and Wolfe, 1977), which describe a trade-off between perceived benefits and risks. Notably, effect sizes for benefits outweighed those of risks in shaping behaviours regarding AI chatbot use. A possible explanation may be that certain risks of AI use do not directly affect the user, or they may not have experienced the risks yet. Risks such as receiving poor advice may only become apparent after the damage is done. Furthermore, previous work has shown that individuals weigh distant and immediate risks differently and that immediate benefits are often more important than long-term risks (Strathman et al., 1994; Acquisti and Grossklags, 2005). For instance, benefits such as efficiency may outweigh less familiar and long-term risks such as the environmental impact. Future research could explore which benefits and risks are perceived more often or strongly, and how these shape perceptions and behaviours.

Moreover, we studied the difference in perceived benefits and risks, intention to use, and willingness to self-disclose across topic type and sensitivity (RQ1). The intention to use an AI chatbot was higher in low-sensitive conditions compared to high-sensitive conditions. There was no significant association between topic type or sensitivity and our other variables of interest. This in line with previous work who found that chatbots are the least acceptable source of information for (mental) health, although stigma could increase this acceptability slightly (Miles et al., 2021; Liu et al., 2025b),

Despite the high number of AI-users among our sample, more than half had a relatively low intention to use AI chatbots. Willingness to share data in the scenarios was relatively low as well. While topic type and sensitivity did not directly influence the intention and willingness, they might influence the importance attached to benefits and risks through normative perceptions (Nissenbaum, 2004).

Furthermore, individual factors were associated with our variables of interest. Participants with no experience with AI chatbots and lower AI literacy were significantly less likely to use chatbots and self-disclose, compared to higher AI literate participants. This could be explained by the fact that active usage is likely to influence intended usage in our scenarios. As AI chatbots are becoming more widely adopted throughout society, the perspectives may shift over time making it worthwhile to repeat this study. Coping style and experience with health institutions were also key factors: Participants who already search for alternative information (e.g., consulting other providers or searching independently) were also more likely to use an AI chatbot while those who mainly relied on asking questions to their own doctor were less likely to do so. Future research could further examine this link between health information-seeking behaviour and AI chatbot use.
In line with earlier studies where higher educated people perceived less benefits (Platt et al., 2024), we found that people with higher education perceive significantly more risks which in turn could closely related to perceiving less benefits. Unlike previous work, we found no effects of gender, age or general health while previous work did (Platt et al., 2024; Mendel et al., 2025; Sumner et al., 2025). The low intentions and willingness overall might have played a role but also the more widespread awareness and use of AI chatbots might have diminished previously found effects.

## 5.1 Strengths and limitations

Given the large scope of the study, including health contexts that are usually studied in isolation, we were able to draw conclusions that go beyond specific health contexts. Additionally, we were able to further investigate the role of individual characteristics in shaping perceptions and (intended) behaviours because of the large and representative sample. The rapid adoption of AI chatbots for health-related questions makes these insights valuable for researchers who aim to understand decision-making regarding health-related AI chatbot use, as well as for policymakers and developers. Healthcare providers could pay extra attention to people who may not be aware of the possible risks associated with using AI chatbots for health-related questions.

Some limitations of the current study can be highlighted as well. First, even though we were careful to select low-sensitive conditions that were less sensitive than high-sensitive conditions, all psychological conditions were perceived as more sensitive than physical conditions. As also mentioned by Kowalski and Peipert (2019), people perceive physical and psychological conditions different with psychological conditions having a higher score on public stigma, indicating an inherent difference in sensitivity levels given the topic type. Therefore, caution is advised when directly comparing physical and psychological conditions. Moreover, we acknowledge that topic sensitivity can differ between cultures and the topics marked as high-sensitive in this study may not be equally sensitive in other cultures.

Second, participants were asked to imagine having a certain health condition and using an AI chatbot. Not all participants had experience with these conditions, and their responses may differ from actual trade-offs between the benefits and risks in a similar situation in real-life. Contextual factors, such as the time of the day, could also influence this trade-off (Costa-Gomes et al., 2026). Future research could further investigate the role of topic type and sensitivity in real-life contexts, for instance by analysing actual conversations with AI chatbots surrounding different topics people experience in daily life.

### 5.2 Conclusion

Given AI's growing role in health-related information seeking, it is important to reflect on its impact and the factors associated with public perceptions. Our systematic comparison showed that perceived benefits were positively related to intention to use an AI chatbot and self-disclosure, whereas perceived risks were inversely related to these outcomes. While the type of health-condition discussed with the AI chatbot is no important factor, aside from small differences in intentions across topics with differing sensitivity levels, individual characteristics are associated with the perception and behaviour of users. The sensitive nature of health information requires extra attention, and this work provided first insights which can guide future research on informed benefit–risk trade-offs and the appropriate implementation of AI chatbots across different health domains.

## Declaration of conflicting interests

The authors declare no potential conflicts of interest.

## Funding statement

This work was funded by the Dutch Research Council (NWO) Take-Off grant (54002069).

## Ethical approval and informed consent statements

Ethical approval for the study was received from the Research Ethics and Data Management Committee of Tilburg University (2025.46) prior to data collection. All participants provided informed consent through the panel prior to participation.

## Data availability statement

Data related to the project can be found on OSF for the pretest (https://osf.io/njbp5/overview?view_only=7083664909124b4d8389424afd8681d7) and accessed through the LISS Archive for the main study (https://www.dataarchive.lissdata.nl/study-units/view/1) once the data is published online.

# 6 References

Acquisti A and Grossklags J (2005) Privacy and rationality in individual decision making. *IEEE Security & Privacy* 3(1): 26-33.

Antes AL, Burrous S, Sisk BA, et al. (2021) Exploring perceptions of healthcare technologies enabled by artificial intelligence: An online, scenario-based survey. *BMC Medical Informatics and Decision Making* 21(1).

Benda N, Desai P, Reza Z, et al. (2024) Patient perspectives on AI for mental health care: Cross-sectional survey study. *JMIR Mental Health* 11: e58462.

Costa-Gomes B, Tolmachev P, Taysom E, et al. (2026) Public use of a generalist LLM chatbot for health queries. *Nature Health*. DOI: 10.1038/s44360-026-00117-x.

de León E, Wan J, Votta F, et al. (2024) Public values in the algorithmic society longitudinal panel survey. University of Amsterdam.

Haensch A-C (2025) " It listens better than my therapist": Exploring social media discourse on LLMs as mental health tool. *arXiv preprint arXiv:2504.12337*.

Janz NK and Becker MH (1984) The health belief model: A decade later. *Health Education Quarterly* 11(1): 1-47.

Jia X, Pang Y and Liu LS (2021) Online health information seeking behavior: A systematic review. *Healthcare* 9(12): 1740.

Kieskompas (2023) *Het politieke landschap*. Available at: https://tweedekamer2023.kieskompas.nl/nl/results/compass (accessed 07/04/2026).

Kowalski RM and Peipert A (2019) Public- and self-stigma attached to physical versus psychological disabilities. *Stigma and Health* 4(2): 136-142.

Laufer RS and Wolfe M (1977) Privacy as a concept and a social issue: A multidimensional developmental theory. *Journal of Social Issues* 33(3): 22-42.

Liu Z, Zou W and Lin C (2025a) Exploring the influence of privacy concerns, AI literacy, and perceived health stigma on AI chatbot use in healthcare: An uncertainty reduction approach. *Patient Education and Counseling* 140: 109271.

Liu Z, Zou W and Yang S (2025b) What drives patient preferences for AI chatbots over doctors? A survey study using the O-S-O-R model. *Health Communication*. DOI: 10.1080/10410236.2025.2537814. 1-12.

Long D and Magerko B (2020) What is AI literacy? Competencies and design considerations. *Proceedings of the 2020 CHI Conference on Human Factors in Computing Systems.* Association for Computing Machinery, 1-16.

Lythreatis S, Singh SK and El-Kassar A-N (2022) The digital divide: A review and future research agenda. *Technological Forecasting and Social Change* 175: 121359.

Macleod A, Cameron P, Geringer J, et al. (2025) How to … study sensitive topics. *The Clinical Teacher* 22(5).

McBain RK, Cantor JH, Breslau J, et al. (2026) AI chatbot use and disclosure for mental health among US adolescents and young adults. *JAMA Pediatrics*. DOI: 10.1001/jamapediatrics.2026.2015.

Mendel T, Singh N, Mann DM, et al. (2025) Laypeople's use of and attitudes toward large language models and search engines for health queries: Survey study. *Journal of Medical Internet Research* 27: e64290.

Meng X and Liu J (2025) "Talk to me, i'm secure": Investigating information disclosure to AI chatbots in the context of privacy calculus. *Online Information Review* 49(5): 933-954.

Menon D and Shilpa K (2023) "Chatting with ChatGPT": Analyzing the factors influencing users' intention to use the open ai's ChatGPT using the UTAUT model. *Heliyon* 9(11): e20962.

Miles O, West R and Nadarzynski T (2021) Health chatbots acceptability moderated by perceived stigma and severity: A cross-sectional survey. *DIGITAL HEALTH* 7: 205520762110630.

Nissenbaum H (2004) Privacy as contextual integrity. *Washington Law Review* 79: 119.

Norman CD and Skinner HA (2006) eHEALS: The eHealth literacy scale. *Journal of Medical Internet Research* 8(4): e507.

Nyakhar S and Wang H (2025) Effectiveness of artificial intelligence chatbots on mental health & well-being in college students: A rapid systematic review. *Frontiers in Psychiatry* 16.

Omarzu J (2000) A disclosure decision model: Determining how and when individuals will self-disclose. *Personality and Social Psychology Review* 4(2): 174-185.

Platt J, Nong P, Smiddy R, et al. (2024) Public comfort with the use of ChatGPT and expectations for healthcare. *Journal of the American Medical Informatics Association* 31(9): 1976-1982.

R Core Team (2016) R: A language and environment for statistical computing. Vienna, Austria: R Foundation for Statistical Computing.

Reeskens T and Metz A (2023) Personal values and Covid-19. In: Centerdata (ed). Centerdata.

Shaban M, Osman YM, Mohamed NA, et al. (2025) Empowering breast cancer clients through AI chatbots: Transforming knowledge and attitudes for enhanced nursing care. *BMC Nursing* 24(1).

Singg S and Pena G (2025) AI chatbots in medical and mental healthcare: A provider-focused review of benefits, risks, and applications. *Journal of Clinical Review & Case Reports* 10(12): 01-06.

Singh B, Olds T, Brinsley J, et al. (2023) Systematic review and meta-analysis of the effectiveness of chatbots on lifestyle behaviours. *NPJ Digital Medicine* 6(1).

Statistics Netherlands *Standaard onderwijsindeling 2021*. Available at: https://www.cbs.nl/nl-nl/onze-diensten/methoden/classificaties/onderwijs-en-beroepen/standaard-onderwijsindeling--soi--/standaard-onderwijsindeling-2021 (accessed 19/05/2026).

Strathman A, Gleicher F, Boninger DS, et al. (1994) The consideration of future consequences: Weighing immediate and distant outcomes of behavior. *Journal of Personality and Social Psychology* 66(4): 742-752.

Sumner J, Wang Y, Tan SY, et al. (2025) Perspectives and experiences with large language models in health care: Survey study. *Journal of Medical Internet Research* 27: e67383.

Tang H, Ou M and Zheng H (2025) Between promise and peril: Users' risk–benefit trade-offs in their generative AI usage. *Big Data & Society* 12(4).

Taylor DA and Altman I (1975) Self-disclosure as a function of reward-cost outcomes. *Sociometry*. DOI: 10.2307/2786231. 18-31.

Van Der Vaart R, Van Deursen AJ, Drossaert CH, et al. (2011) Does the eHealth literacy scale (eHEALS) measure what it intends to measure? Validation of a dutch version of the eHEALS in two adult populations. *Journal of Medical Internet Research* 13(4): e86.
Van Zuuren FJ, De Groot KI, Mulder NL, et al. (1996) Coping with medical threat: An evaluation of the Threatening Medical Situations Inventory (TMSI). *Personality and Individual Differences* 21(1): 21-31.
Venkatesh V, Morris MG, Davis GB, et al. (2003) User acceptance of information technology: Toward a unified view. *MIS Quarterly* 27(3): 425-478.
Wang B, Rau P-LP and Yuan T (2022) Measuring user competence in using artificial intelligence: Validity and reliability of artificial intelligence literacy scale. *Behaviour & Information Technology* 42(9): 1324-1337.
Weidinger L, Uesato J, Rauh M, et al. (2022) Taxonomy of risks posed by language models. *Proceedings of the 2022 ACM Conference on Fairness, Accountability, and Transparency.* Association for Computing Machinery, 214-229.
Xiao Y, Zhou KZ, Liang Y, et al. (2024) Understanding the concerns and choices of public when using large language models for healthcare. *arXiv preprint arXiv:2401.09090*.
Yi Y, Wu Z and Tung LL (2005) How individual differences influence technology usage behavior? Toward an integrated framework. *Journal of Computer Information Systems* 46(2): 52-63.
Yun HS and Bickmore T (2025) Online health information–seeking in the era of large language models: Cross-sectional web-based survey study. *Journal of Medical Internet Research* 27: e68560.